\documentclass[aps,prl,twocolumn,amsmath,amssymb,nofootinbib,superscriptaddress,floatfix,reprint,longbibliography]{revtex4-1}
\usepackage[T1]{fontenc}
\usepackage{amsmath}
\usepackage{amsfonts}
\usepackage{amsthm}
\usepackage{amssymb}
\usepackage{lipsum}
\usepackage{graphicx}
\usepackage[usenames,dvipsnames]{color}
\usepackage{float}
\usepackage{multirow}
\usepackage{soul}
\usepackage{appendix}
\usepackage[colorlinks=true, urlcolor=blue, linkcolor=blue, citecolor=blue, pdftex]{hyperref}
\usepackage{dcolumn}
\usepackage{bm}
\usepackage{url}
\usepackage{braket}
\usepackage[range-units=single,separate-uncertainty]{siunitx}
\usepackage{verbatim} 
\usepackage{booktabs}
\usepackage[version=4]{mhchem}

\begin{document}
\author{Savita Chaudhary}
\affiliation{Department of Physical Sciences, Indian Institute of Science Education and Research (IISER) Mohali, Knowledge City, Sector 81, Mohali 140306, India.}

\author{Shama}
\affiliation{Department of Physical Sciences, Indian Institute of Science Education and Research (IISER) Mohali, Knowledge City, Sector 81, Mohali 140306, India.}

\author{Jaskaran Singh}
\affiliation{Department of Physical Sciences, Indian Institute of Science Education and Research (IISER) Mohali, Knowledge City, Sector 81, Mohali 140306, India.}
\affiliation{Department of Physics, Punjabi University, Patiala, 147002, India.}

\author{Armando Consiglio}\affiliation{Institut f\"{u}r Theoretische Physik und Astrophysik and W\"{u}rzburg-Dresden Cluster of Excellence ct.qmat, Universit\"{a}t W\"{u}rzburg, 97074 W\"{u}rzburg, Germany}

\author{Domenico Di Sante}\affiliation{Department of Physics and Astronomy, Alma Mater Studiorum, University of Bologna, 40127 Bologna, Italy}\affiliation{Center for Computational Quantum Physics, Flatiron Institute, 162 5th Avenue, New York, New York 10010, USA}

\author{Ronny Thomale}\affiliation{Institut f\"{u}r Theoretische Physik und Astrophysik and W\"{u}rzburg-Dresden Cluster of Excellence ct.qmat, Universit\"{a}t W\"{u}rzburg, 97074 W\"{u}rzburg, Germany}

\author{Yogesh Singh}
\affiliation{Department of Physical Sciences, Indian Institute of Science Education and Research (IISER) Mohali, Knowledge City, Sector 81, Mohali 140306, India.}

\date{\today}

\title{Role of electronic correlations in the Kagome lattice superconductor LaRh$_3$B$_2$}

\begin{abstract}
LaRh$_3$B$_2$ crystallizes in a layered structure where Rh atoms form a perfect Kagome lattice.  The material shows superconductivity at $T_c \approx 2.6$~K\@ and no signature for density wave instabilities. We report our measurements of electronic transport, magnetization, and heat capacity in the normal and superconducting state, and derive normal and superconducting parameters. From first principles calculations of the electronic band structure, we identify all features of Kagome bands predominantly formed by the Rh $d$ orbitals: a flat band, Dirac cones, and van Hove singularities. The calculation of the phonon dispersions and electron-phonon coupling suggests a strong similarity between LaRh$_3$B$_2$ and AV$_3$Sb$_5$ (A=K,Cs,Rb). For LaRh$_3$B$_2$, it matches quantitatively with the observed $T_c$, supporting a conventional phonon mediated pairing mechanism.  By comparison to the $A$V$_3$Sb$_5$ family, we conjecture a reduced importance of electron correlations in LaRh$_3$B$_2$. 
\end{abstract}

\maketitle 
\section{Introduction} 
The kagome lattice has long been a playground for novel physics in condensed matter. Insulating kagome lattice realizations with localized magnetic moments are platforms to explore the effects of geometric magnetic frustration.  The quantum spin liquid (QSL) ground state in the mineral Herbertsmithite ZnCu$_3$(OH)Cl$_2$ is a prime example of this behaviour \cite{Fu2015, Han2012}.  Insulating quantum magnets with a kagome network in higher dimensions have also shown novel frustrated magnetism and QSL behaviour \cite{Balz, Okamoto2007, Singh2013}. More recently metallic kagome lattice materials have been brought into focus due to the prediction that the electronic structure of electrons on a kagome lattice might allow to access correlated Dirac cones or van Hove singularities near the Fermi energy\cite{Kiesel2013,Wang2013,Mazin2014}.  The two-dimensional kagome lattice has features in its band structure which provide, even still in the itinerant limit, the opportunity of marrying non-trivial topology and strong electron correlations.  Search for realizations of a material with an ideal isolated kagome lattice is therefore a fundamentally important quest.  In recent years a few families of metallic materials possessing a kagome lattice have indeed been reported or theoretically predicted.  These include the Herbertsmithite related material Ga/ScCu$_3$(OH)$_6$Cl$_2$ \cite{Mazin2014}, the magnetic kagome metals CoSn and FeSn \cite{Mingu2020,Ming2020}, and the ferromagnetic kagome metal YMn$_6$Sn$_6$ \cite{Li2021}.  Most recently, the $A$V$_3$Sb$_5$ ($A =~$K, Rb, Cs) family of materials have been discovered and shown to host a perfect kagome network of V ions \cite{Ortiz2019,Neupert2022}.  Evidence for electron correlations and non trivial topology in these materials emerges from the discovery of charge density waves, superconductivity, anomalous Hall effect, and multiple van Hove singularities nearby the Fermi energy \cite{Ortiz2019, Ortiz2020, Ortiz2021,Yang2020}.  

Another family of materials possessing the kagome lattice is $R$T$_3X_2$ ($R =$~Lanthanide, $T =$~ $4$d or $5$d transition metal, $X =$~ Si, B).  These materials were discovered in the 1980s \cite{Ku1980, Barz, Vandenberg} and several of them were reported to show superconductivity with $T_{\rm c}$s between $1$~K to $\sim 7$~K \cite{Ku1980, Barz, Vandenberg, Malik, Athreya, Rauchschwalbe}.  However, most of these studies were not made in the context of the connection of properties with the underlying kagome lattice. Only recently LaRu$_3$Si$_2$, which has the highest $T_{\rm c} = 7$~K in this family of materials, has been studied in relation to the kagome lattice, and several unconventional properties have been reported possibly arising from electron correlations from the flat bands \cite{Li2011, Li2012,Li2016,Mielke2021}.  Materials in the $R$T$_3X_2$ family thus form another promising platform to study the kagome related features in the band structure, and their interplay with superconductivity.  

We report on the electronic structure, phonon profile, and superconducting properties of LaRh$_3$B$_2$ which has previously been reported to show superconductivity at low temperatures, where the reported superconducting $T_{\rm c}$ ranges from $< 1.2$ to $2.8$~K \cite{Ku1980,Malik}.  Our electronic band structure calculations reveal a flat band above the Fermi energy, and van Hove singularities and Dirac cones at several locations in the Brillouin zone including close to the Fermi energy $E_{\rm F}$.  We find that the $E_{\rm F}$ is located at the top of a sharp peak in the density of states (DOS). We use this to address the extreme sample dependence of the superconducting $T_{\rm c}$.  The superconductivity is found to be of conventional weak coupling type.  This is supported by estimations of the $T_c$ from phonon calculations and the estimate of electron-phonon coupling.  The van Hove singularities in the band structure are found to be located a few eVs away from $E_{\rm F}$, which is large against the characteristic ordering scales and thus explains why these materials do not show signals of correlation-induced phenomena such as  charge density waves or other instabilities.  This is also supported by the phonon calculations which stress the absence of any imaginary frequency mode.  In addition, we observe anomalous temperature dependencies of the magnetic susceptibility and heat capacity, and a slightly enhanced Sommerfeld coefficient,  which we argue to arise from the narrow band which is part of the DOS.  In comparison with the $A$V$_3$Sb$_5$ materials, our results point to a reduced importance of electronic-correlations in the LaRh$_3$B$_2$ material.

\section{Methods}
Polycrystalline samples of LaRh$_3$B$_2$ were synthesized by arc-melting stoichiometric ratios of La (3N, Alfa Aesar), Rh (5N, Alfa Aesar) and B (6N, Alfa Aesar). The melted buttons were flipped over and melted $5$--$10$ times to promote homogeneity.  Powder X-ray diffraction (PXRD) on a Bruker D8 Advance diffractometer system with Cu-K$\alpha$ radiation was used to determine the phase purity of the arc-melted LaRh$_3$B$_2$ sample. The relative stoichiometry of La and Rh was confirmed using energy dispersive spectroscopy using a scanning electron microscope. The dc magnetic susceptibility $\chi$, heat capacity $C$, and electrical transport were measured using a Quantum Design Physical Property Measurement System equipped with a He3 insert.   To theoretically simulate the electronic structure of LaRh$_3$B$_2$, we performed first-principles
density functional theory (DFT) calculations using the Vienna Ab initio simulation package
(VASP) \cite{Kresse93,Kresse94,Kresse96a,Kresse96b}. We considered the projector-augmented wave (PAW) pseudo potential with
exchange-correlation functional of generalized gradient approximation (GGA) of Perdew-Burke-Ernzerhof \cite{Kresse,Perdew}. Starting with the experimental structure, the lattice relaxation was performed to optimize the crystal structure by using variable cell relaxation. We adopted a $12\times12\times12$ k mesh for the first Brillouin zone. We have used an energy cut-off of $450$~eV for the plane wave basis. The convergence criteria for energy and force are set to $10^{-6}$~eV and $0.02$~eV/\AA, respectively. In the DFT calculation, spin-orbit coupling was not included. However, we have used scalar relativistic potential which takes scalar relativistic effects into account.  Phonon calculations have been performed using density functional perturbation theory, as implemented in Quantum Espresso \cite{Giannozzi2020, Giannozzi2009, Giannozzi}. 
 Exchange and correlation effects were included using the generalized gradient approximation (GGA) with the Perdew-Burke-Ernzerhof (PBE) functional \cite{Perdew}; the pseudopotentials are norm-conserving, with core correction, and scalar relativistic \cite{Hamann}.\\
Self-consistent calculations of the previously relaxed unit cell have been performed with a 8$\times$8$\times$12 $k$-grid. The kinetic energy cutoff for the wavefunctions is equal to 100 Ry, while the cutoff for charge density is 400 Ry.\\
Convergence threshold for ionic minimization and electronic self-consistency are set to be 1.0D-15.\\
The self-consistency threshold for phonon calculations is 1.0D-15 as well, with a $q$-grid of 4$\times$4$\times$2.\\
Non-self consistent calculations for the density of states have been performed with a 60$\times$60$\times$48 $k$-grid.\\
Finally, the electron-phonon interaction is computed via an interpolation over the Brillouin Zone \cite{Wierzbowska}.

\begin{figure}[t]   
\includegraphics[height = 2.4 in, width=3.25 in]{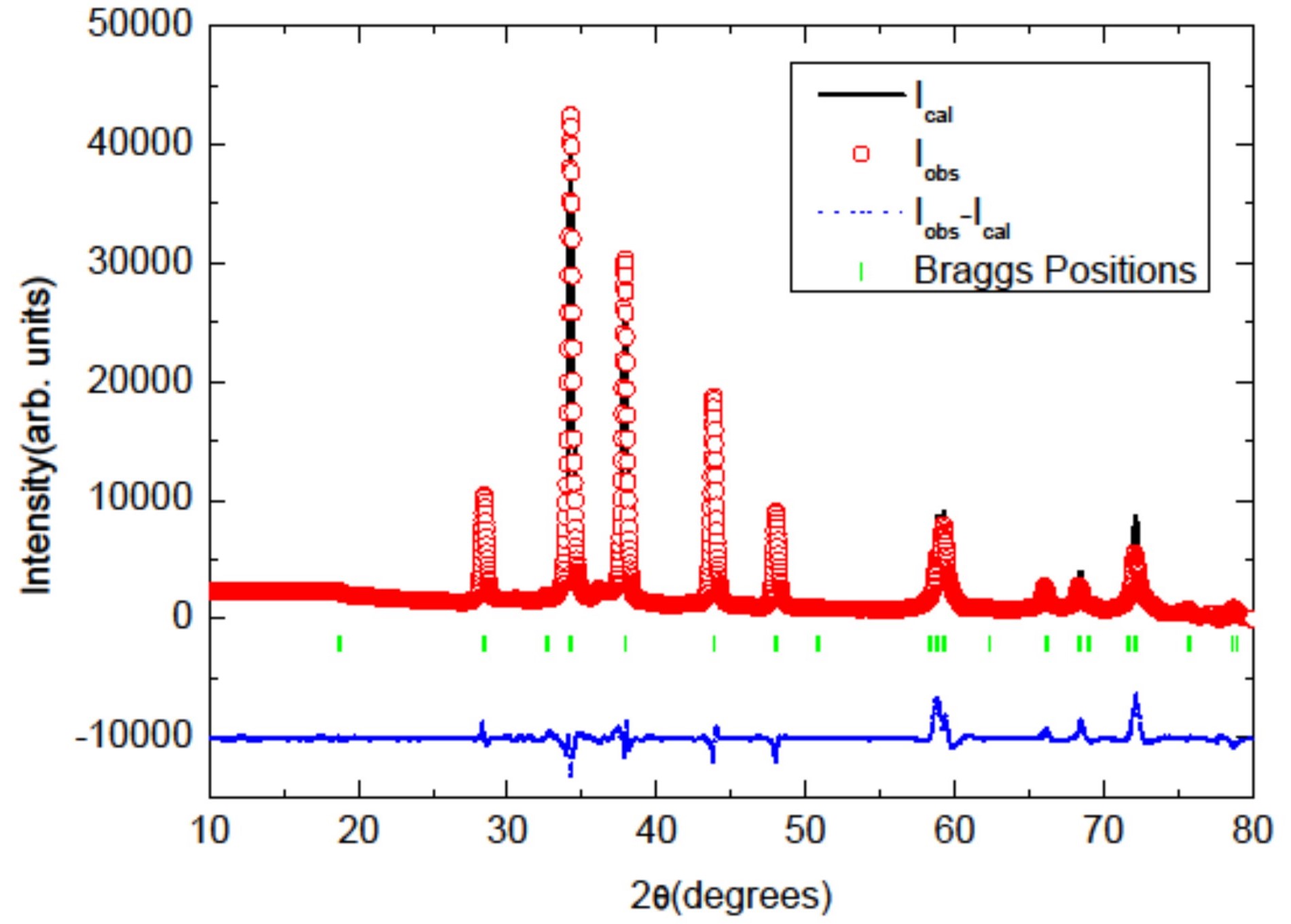}
\caption{(Color online) Powder x-ray diffraction and results of refinement. 
\label{Fig-xrd}}
\end{figure}

\begin{figure*}[t]   
\includegraphics[width= 7 in]{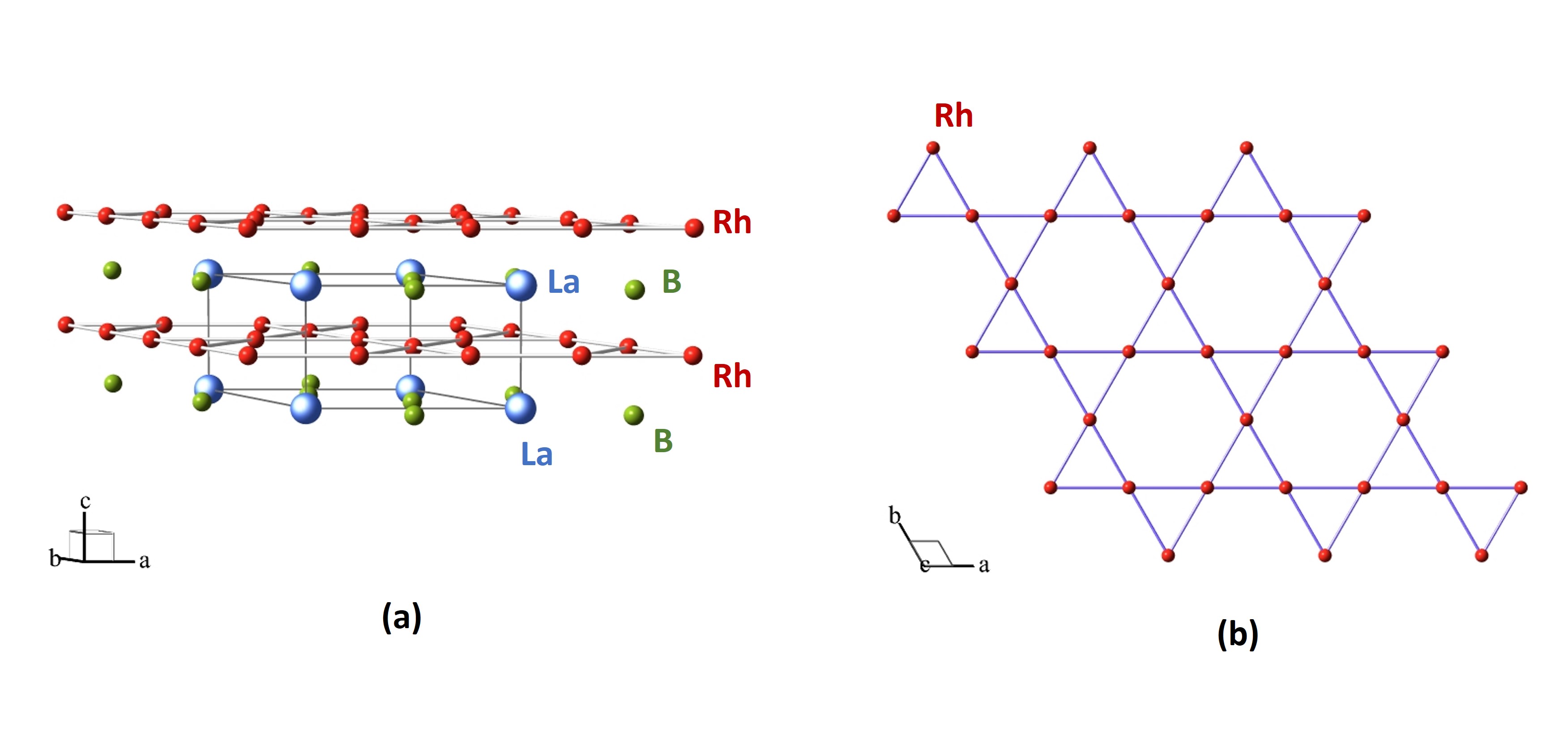}
\caption{(Color online) (a) A schematic of the crystal structure of LaRh$_3$B$_2$ viewed perpendicular to the crystallographic $c$-axis showing the layered nature of the structure with Rh atomic planes separated along the $c$-axis by planes made up of La and B atoms.  (b) Viewed along the $c$-axis, the Rh atoms form an undistorted kagome lattice. 
\label{Fig-structure}}
\end{figure*}

\section{Structure}
LaRh$_3$B$_2$ crystallizes in a honeycomb structure with space group $P6/mmm$. There are no variable parameters in the structure apart from the unit cell size.  The powder x-ray diffraction (PXRD) is shown in Fig.~\ref{Fig-xrd}. The PXRD pattern confirmed that the synthesized material is single phase and a refinement, shown in Fig.~\ref{Fig-xrd}, of the powder pattern gave lattice parameters $a = 5.486$\AA~ and $c = 3.136$\AA.  In literature, a range of values for the lattice parameters have been reported and our values fall within this range of values \cite{Ku1980}.  We will make a connection of unit cell parameters with the electronic properties later.  A schematic of the crystal structure of LaRh$_3$B$_2$ is shown in Fig.~\ref{Fig-structure}.  The structure is made up of layers of Rh planes separated by planes of La and B stacked along the $c$-axis, as shown in Fig.~\ref{Fig-structure}(a).  The arrangement of the Rh atoms within the Rh-planes is a perfect kagome lattice as shown in Fig.~\ref{Fig-structure}(b).  These materials therefore have the structural ingredients to show electronic structure features expected for a kagome metal.  It must be noted however, that the short $c$-axis necessarily means that coupling between kagome planes may be significant.

\section{Results}
\subsection{Electronic Band Structure}
\begin{figure*}[t]   
\includegraphics[width = 7in]{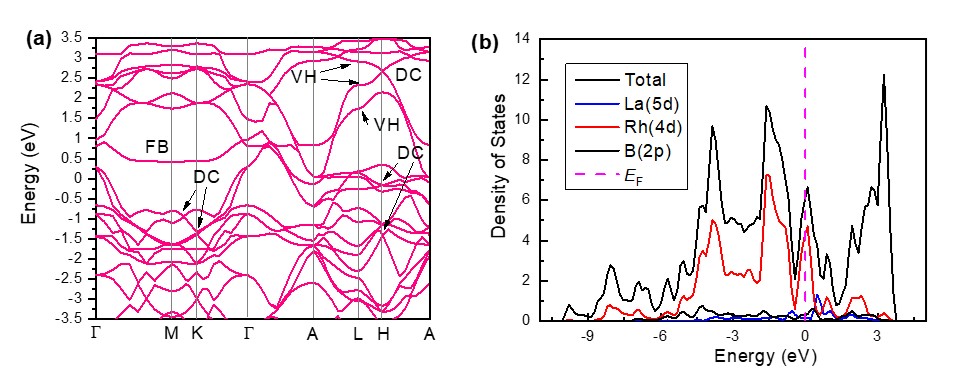}    
\caption{(Color online) (a) The electronic structure of LaRh$_3$B$_2$ along high symmetry directions in $k$-space.  The boxes are to highlight some interesting features as discussed in the main text  (b) The total and partial density of states as a function of energy measured from the Fermi energy. 
\label{Fig-dos}}
\end{figure*}

Figure~\ref{Fig-dos}(a) shows the electronic band structure for LaRh$_3$B$_2$ along some high symmetry directions in the Brillouin zone.  It is evident that several bands cross the Fermi level, confirming that LaRh$_3$B$_2$ is a metal.  The total and partial density of states (DOS) are shown in Fig.~\ref{Fig-dos}(b).  The Fermi level ($E_F$) is situated near the top of a very narrow band resulting in a fairly large DOS at $E_F$ of $6.6$~states/eV\@.  From the partial DOS it is clear that the majority contribution to the total DOS comes from Rh $4$d orbitals and both La and B contribute very small amounts to the total DOS at $E_F$. The narrow band at $E_F$ leads to a strong sensitivity of the superconducting $T_c$ on the unit cell size and to other anomalous physical properties as we will discuss later.  

We now turn to the novel features of the band structure arising from the kagome Rh planes.  As can be seen in Fig.~\ref{Fig-dos}(a), we observe a flat band (FB) in the $\Gamma - M - K - \Gamma$ direction about $0.4$~eV above $E_F$.  This flat band is separate from any other bands.  Another series of disconnected flat bands are observed along the $\Gamma - A$ direction about $0.75$~eV above $E_F$.  In addition to these flat portions of the electronic dispersion, Dirac cones (DC) are observed at several locations in the band structure.  There are Dirac bands $140$~meV below and $2.75$~eV above $E_F$ at $H$ and a Dirac cone about $1$~eV below $E_F$ along the $M - K$ direction in the BZ.  We also identify van Hove (VH) singularities located symmetrically above and below the Dirac cone at 2.75~eV.  Thus the band structure of LaRh$_3$B$_2$ possesses the predicted features of the kagome lattice band structure near $E_F$ with modifications arising most likely from the three-dimensional nature of the material.  

\subsection{Physical Properties}
\begin{figure*}[t]   
\includegraphics[width = 7.35in]{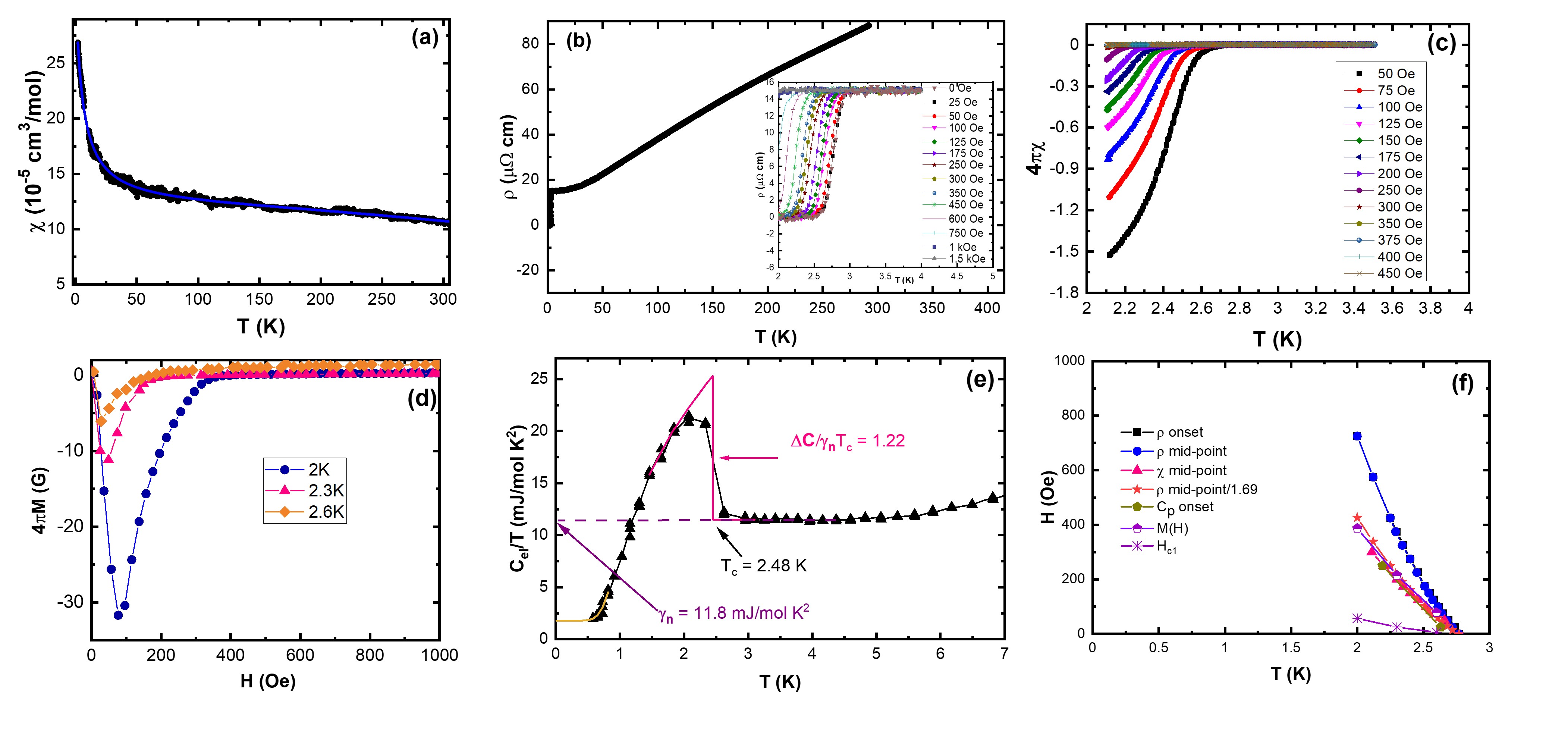}    
\caption{(Color online) (a) The normal state magnetic susceptibility at $H = 2$~T, (b) Resistivity versus temperature at zero field.  Inset shows the superconducting transition at various fields, (c) Dimensionless magnetic susceptibility ($4\pi\chi$) in the superconducting state at various fields, and (d) the electronic contribution to the zero field specific heat $C_{\rm el}/T$ versus $T$.
\label{Fig-LRB}}
\end{figure*}

Figure~\ref{Fig-LRB} shows the electrical, magnetic, and thermal properties of LaRh$_3$B$_2$ in the normal and superconducting states.  Figure~\ref{Fig-LRB}~(a) shows the magnetic susceptibility $\chi$ versus temperature $T$ between $2$~K and $300$~K in an applied magnetic field of $H = 2$~T\@.  At low temperatures, small amounts of magnetic impurities lead to a Curie like upturn. The $\chi$ is found to be temperature dependent in the whole temperature range. This is not what is  expected for a Pauli paramagnetic metal where a $T$ independent $\chi$ is expected.  This $T$ dependent $\chi$ arises due to the $E_F$ being situated on a narrow peak in the DOS.  The change in temperature results in a change in the DOS at $E_F$ leading to a $T$ dependent Pauli paramagnetic susceptibility.  To support this idea the $\chi(T)$ in the full temperature range was fit with the expression $\chi(T) = \chi_o[1-(T/T_E)^2]+C/(T-\theta)$, where the first term represents the $T$ dependent Pauli paramagnetic susceptibility and the second term represents the contribution from the small amounts of magnetic impurities which give rise to the Curie like upturn in $\chi(T)$ at the lowest temperatures.  The fitting parameters are $\chi_o$ the temperature independent average Pauli paramagnetic susceptibility, $T_E$ which is a phenomenological parameter related to the Fermi energy, $C$ which is the Curie constant of the impurities, and $\theta$ which is the Weiss temperature representing any interactions between the magnetic impurities.  A very good fit with the above expression was obtained and is shown as the solid curve through the data in Fig.~\ref{Fig-LRB}~(a).  The fit parameters obtained were $\chi_o = 11.8(2)\times 10^{-5}$~G~cm$^3$/mol, $T_E = 860(7)$~K, $C = 0.0010(4)$~G~cm$^3$~K/mol, and $\theta = -5.5(1)$~K\@.  This value of  $C$ is equivalent to $0.25\%$ of $S = 1/2$ impurities, which is quite small. 

Figure~\ref{Fig-LRB}~(b) shows the electrical resistivity $\rho$ in zero field between $2$~K and $300$~K\@. We observe metallic behaviour with a residual resistivity ratio RRR $= \rho(300{\rm K})/\rho(2{\rm K}) \approx 5$.  
The inset in Fig.~\ref{Fig-LRB}~(b) shows the $\rho(T)$ data below $T = 4$~K measured in various applied fields.  The sharp drop to zero resistance below $T_c \sim 2.6$~K in zero field signals the onset of superconductivity in LaRh$_3$B$_2$.  Further evidence of a superconducting state is obtained from the diamagnetism observed in magnetic measurements shown in Fig.~\ref{Fig-LRB}~(c) and (d).  From Fig.~\ref{Fig-LRB}~(c) we can see that the value of $4\pi\chi$ in the superconducting state is greater than $-1$ suggesting demagnetization factors due to the irregular shape of the sample. The magnetization data in Fig.~\ref{Fig-LRB}~(d) show behaviour typical of a Type-II superconductor.  The bulk nature of superconductivity is confirmed from heat capacity measurements.  Figure~\ref{Fig-LRB}~(e) shows the electronic heat capacity $C_{\rm el}$ divided by $T$ versus $T$.  The $C_{\rm el}$ is obtained by subtracting a lattice term ($\sim T^3$) from the total heat capacity.  It is of interest to note that the $C_{\rm el}/T$ in the normal state would be expected to be $T$ independent.  However, this is true only for $T \leq 4.5$~K while there is a strong $T$ dependence of $C_{\rm el}/T$ above these temperatures.  This suggests that the lattice contribution may not just have a $T^3$ term but an-harmonic terms may also contribute to the lattice heat capacity.    A sharp anomaly at the onset of the superconducting transition can clearly be seen in Fig.~\ref{Fig-LRB}~(e).  To evaluate the magnitude of the jump in heat capacity at the transition and to obtain an alternate estimate of the bulk superconducting $T_c$, we use an equal entropy construction.  The result is shown as the solid curve through the data near $T_c$.  This gives a value $T_c \approx 2.5$~K\@.  The normal state data above $T_c$ can be extrapolated to $T = 0$ to give an estimate of the Sommerfeld coefficient $\gamma_{\rm n} = 11.8$~mJ/mol~K$^2$.  With this we obtain the jump height at $T_c$ to be $\Delta C_{\rm el}/\gamma_{\rm n} T_c \approx 1.2$, which is smaller than the value $1.43$ expected for a weak coupling single gap superconductor.  The $C_{\rm el}$ data at the lowest temperatures seem to extrapolate to a finite value suggesting some residual contribution from normal electrons. 
The $C_{\rm el}$ data below $T = 0.8$~K were fit by the expression $C_{\rm el} = \gamma_{\rm res}T+$ Aexp$^{-\Delta/k_BT}$ assuming a fully gapped superconducting state.  The $\gamma_{\rm res}T$ term represents the contribution from any non-superconducting fraction of electrons.  An excellent fit, shown in Fig.~\ref{Fig-LRB}~(e), was obtained with the following values for the parameters $\gamma_{\rm res} = 1.7$~mJ/mol~K$^2$ and $\Delta = 6$~K\@.  This value of $\gamma_{\rm res}$ suggests that $\approx 14\%$ electrons do not participate in the superconductivity.  So we must revise our estimate of $\Delta C_{\rm el}/\gamma T_c$ using $\gamma = \gamma_n - \gamma_{\rm res}$.  This gives the value $\Delta C_{\rm el}/\gamma T_c \approx 1.44$ which is close to the value expected from a weak coupling single gap superconductor.  These results suggest that LaRh$_3$B$_2$ is a weak coupling single-gap Type-II superconductor.

From various measurements in finite magnetic field we can track the $T_c$ as a function of the field $H$.  The $H$-$T$ diagram obtained from the various measurements is shown in Fig.~\ref{Fig-LRB}~(f) where both the lower critical field $H_{C1}$ and the upper critical field $H_{C2}$ are shown.  We observe that the $H_{C2}$ data from all measurements except the resistivity measurements agree with each other, while the critical field measured from $\rho$ are consistently higher than values measured from other bulk probes like magnetization and heat capacity.  Such observations have been reported previously in some materials and have been linked to surface superconductivity \cite{Zeinali,Pradip}.  It has been shown that the critical field for surface superconductivity is $\approx 1.69 H_{c2}$, where $H_{c2}$ is the bulk critical field.  We also plot in Fig.~\ref{Fig-LRB}~(f) the $H_{C2}$ obtained from resistivity measurements divided by $1.69$.  The critical field so obtained matches the critical field values obtained from other bulk measurements.  So we will treat the scaled critical field from resistivity measurements as the true bulk critical field $H_{c2}$.  The $H_{c2}$ vs $T$ plot shows an upward curvature in the whole temperature range.  This is unusual and inconsistent with observations for most conventional superconductors.  

To learn about the strength of the electron-phonon coupling we make an estimate of the electron-phonon coupling constant $\lambda_{ep}$ using McMillan's formula, which relates the superconducting transition temperature $T_c$ to $\lambda_{ep}$, the Debye temperature $\theta_D$, and the Coulomb repulsion constant $\mu^{*}$
    \[
        T_c=\frac{\Theta_D}{1.45} \exp \left[-\frac{1.04\left(1+\lambda_{\mathrm{ep}}\right)}{\lambda_{\mathrm{ep}}-\mu^*\left(1+0.62 \lambda_{\mathrm{ep}}\right)}\right]
    \]
which can be inverted to give $\lambda_{ep}$ in terms of $T_c$, $\theta_D$ and $\mu^{*}$ as
   \[
       \lambda_{\mathrm{ep}}=\frac{1.04+\mu^* \ln \left(\frac{\Theta_D}{1.45 T_c}\right)}{\left(1-0.62 \mu^*\right) \ln \left(\frac{\Theta_D}{1.45 T_c}\right)-1.04}
    \]

From the heat capacity measurements, we had obtained $\theta_D$ = 518 K and using $T_c$ = 2.5 K, we get $\lambda_{ep} = 0.43$ and 0.52 for $\mu^{*}=0.10$ and 0.15, respectively.  These values of $\lambda_{ep}$ suggest moderate electron-phonon coupling in LaRh$_3$B$_2$.  This is supported by the estimates of the $\lambda_{ep}$ made from our phonon calculations which will be discussed later.

We now present our estimation of various superconducting parameters using expressions previously collected in Refs.~\onlinecite{Singh2007,Singh2010}.  An estimate of the $T = 0$ upper critical field $H_{c2}(0)$ was made by first making a linear extrapolation of the data near $T_c$ to give the slope ${dH_{c2} \over dT}|_{T_c} = -511$~Oe/K\@. This linear slope can then be used to get an estimate of $H_{c2}(0)$ using the Werthamer-Helfand-Hohenberg (WHH) formula for the clean limit $H_{c2}(0) = -0.693T_c{dH_{c2}\over dT}|_{T_c} = 920$Oe.  From the value of $H_{c2}$ we can now estimate the value of the coherence length $\xi$ by the expression $H_{c2} = \phi_0/2\pi\xi^2$, where $\phi_0 = hc/2e = 2.068
\times 10^{-7}$~G~cm$^2$ is the flux quantum.  Using $H_{c2}(0) = 920$~Oe obtained above, we estimate $\xi(0) = 60$~nm. At $T = 2.3$~K near $T_c$ where $H_{c2} = 250$~Oe we get $\xi(0) = 114$~nm.  We have collected the various normal and superconducting state parameters in Table~\ref{T_TC}.

\begin{table}
\begin{center}
\caption{Normal and superconducting state parameters for  LaRh$_3$B$_2$. Here $\gamma$ is the Sommerfeld coefficient, $\beta$ is the coefficient of the $T^3$ term in the low temperature heat capacity, $\theta_{\rm D}$ is the Debye temperature, $n$ is the electron density, $\xi$ is the superconducting coherence length, $\lambda$ is the penetration depth, $l$ is the electron mean free path, and $v_{\rm F}$ is the Fermi velocity.}
\vspace{0.5cm}
\label{T_TC}
\begin{tabular}{|c|c|}
\hline
\hline
RRR& $\approx 5$ \\ \hline
$\gamma$~(mJ/mol~K$^2$) & 11.8 \\ \hline
$\beta$~(mJ/mol~K$^4$) & 0.084 \\ \hline
$\theta_{\rm D}$~(K) & 520 \\ \hline
$n$~(cm$^{-3})$ & $0.97 \times 10^{23}$ \\ \hline
{T$_{\rm C}$(K)} & {2.6} \\ \hline
$\xi_{2{\rm K}}$(nm)& 114\\ \hline
$\xi_{0{\rm K}}$(nm) &  60 \\ \hline
$\lambda_{0{\rm K}}$(nm) & 5.4 \\ \hline
$l_{4{\rm K}}$(nm) & 43 \\ \hline
$v_{{\rm F}}$(cm/s) & $3.5 \times 10^{8}$ \\ \hline
\hline 
\end{tabular}
\end{center}
\end{table}    

\begin{figure}[t]   
\includegraphics[width = 4.in]{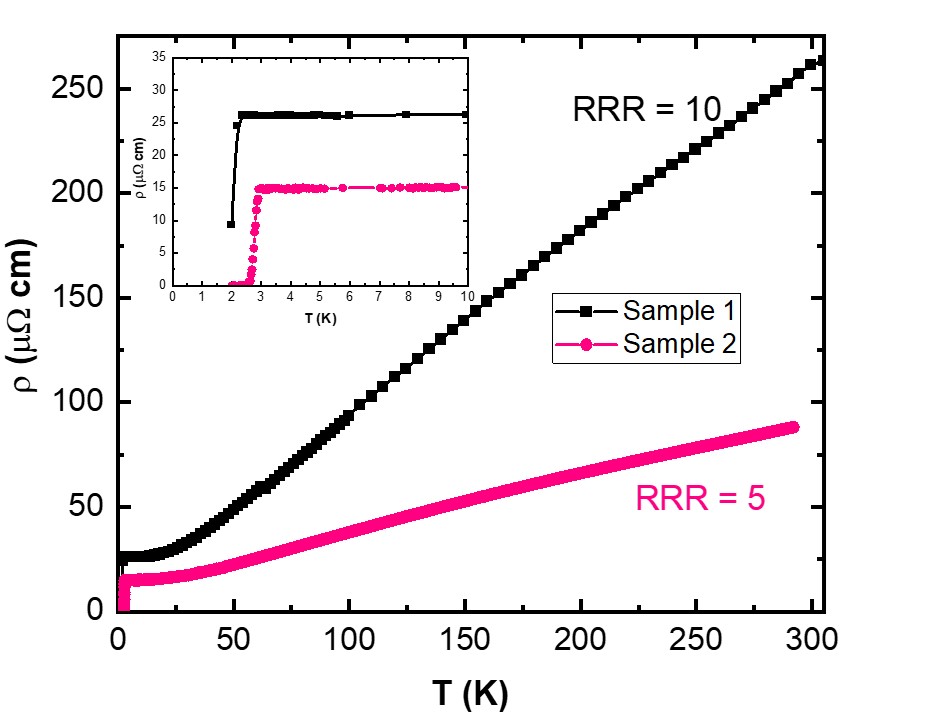}    
\caption{(Color online) Resistivity versus temperature at zero field for two LaRh$_3$B$_2$ samples.  Inset shows the variation in the superconducting transition temperature.
\label{Fig-anne}}
\end{figure}

We now address the large variation in the $T_C$ of LaRh$_3$B$_2$ which has been reported in the literature \cite{Ku1980,Malik}. Figure~\ref{Fig-anne} shows the resistivity of two samples of LaRh$_3$B$_2$ prepared with the same nominal ratios of starting materials.  A refinement of their powder x-ray pattern gave lattice parameters that are slightly different.  The lattice parameters for sample 1 (S1) are $a = 5.484$~\AA~ and $c = 3.139$~\AA~ while those for sample 2 (S2) are $a = 5.486$~\AA~ and $c = 3.136$~\AA.  Thus the S1 has slightly smaller in-plane lattice parameters while its $c$-axis is longer indicating that in this sample the kagome planes are shrunk while the separation between the kagome lattice increases.  S2 on the other hand has a larger kagome plane but the planes are separated by a smaller distance along the $c$-axis.  
From Fig.~\ref{Fig-anne} we see that the electrical transport properties are sensitive to these small changes.  S1 has a larger residual resistivity ratio RRR $= 10$ but has a larger residual resistivity $\rho_0 = 26~\mu \Omega$cm.  S2 on the other hand has a smaller RRR $= 5$ but a smaller residual resistivity $\rho_0 = 15~\mu \Omega$cm.  
From the inset it can be seen that while S2 has a superconductivity onset at $T_c = 2.6$~K, for S1 the onset is $T_c = 2.05$~K, more than $0.5$~K lower.  We also find this large variation in $T_{\rm c}$ with unit cell size of LaRh$_3$B$_2$ in previous reports \cite{Ku1980,Malik}.  For example, superconductivity with a $T_{\rm c} = 2.6$~K was reported for a LaRh$_3$B$_2$ sample with lattice parameters $a = 5.480$\AA~and $c = 3.137$~\AA~\cite{Ku1980}, a $T_{\rm c} = 2.2$~K was reported for a sample with  lattice parameters $a = 5.483$\AA, $c = 3.142$~\AA~ while no superconductivity down to $1.2$~K was found for a sample with lattice parameters $a = 5.512$\AA~and $c = 3.115$~\AA~\cite{Malik}.   

We can address this variation in $T_{\rm c}$ for different samples using our DFT calculations.  Our calculations have shown that the $E_{\rm F}$ lies near the top of a narrow band in the DOS.   This sensitivity of $T_{\rm c}$ most likely originates from changes in the DOS at $E_F$ due to small changes in $E_F$ either by pressure effects (as evidenced by difference in unit cell sizes) or due to a difference in the electron densities (by minute variation in the stoichiometry) in the samples.
A slight change in $E_F$ will lead to large changes in the DOS at $E_F$ because the $E_F$ is located on top of a narrow band in the band structure.

\begin{figure*}[tb]
	\includegraphics[width=.90\textwidth, trim={2.2cm  12cm  2.1cm 3cm},clip]{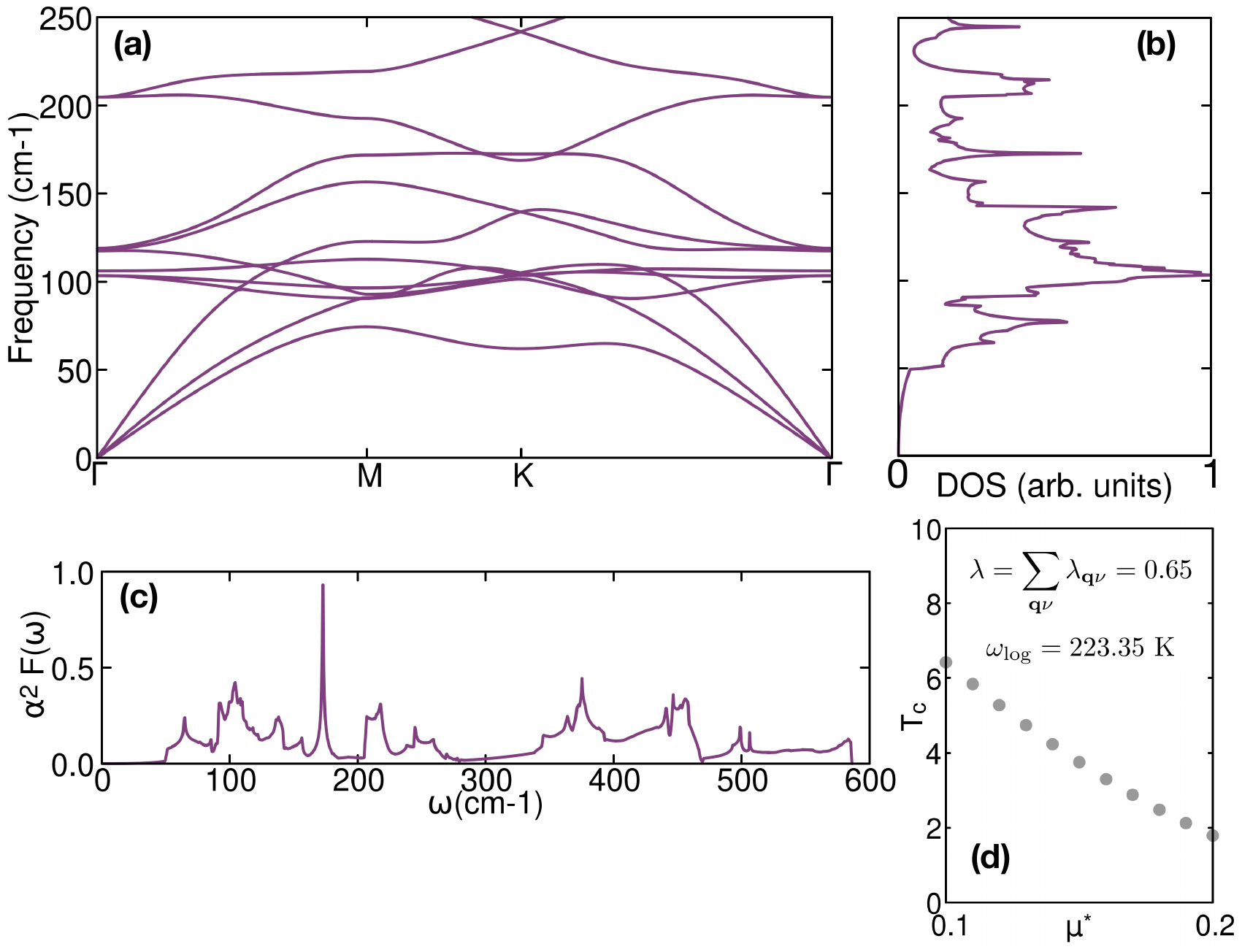}
	\caption{\textbf{a)} Phonon dispersion along high-symmetry lines; \textbf{b)} Corresponding density of states; \textbf{c)} Eliashberg function; \textbf{d)} Superconducting critical temperature $T_c$ as a function of the Coulomb pseudo-potential $\mu^*$.}
	\label{fig:th}
\end{figure*}

Our phonon calculations support the motif of  weak-coupling phonon-mediated superconductivity, and also indicate the reason for the absence of a correlation-induced CDW instability for LaRh$_3$B$_2$.  The phonon dispersion for LaRh$_3$B$_2$ are shown in Figs. \ref{fig:th} a, b.  The phonon dispersion does not exhibit any imaginary frequency mode; this is a sign of dynamical stability, confirming the absence of experimental signatures of charge density wave states.
Low-frequency phonon modes mostly originate from La atoms, while intermediate frequencies are essentially due to Rh atoms; the manifold of low-dispersing bands around 120 cm$^{-1}$ is then due to the kagome network. Finally, B atoms contribute to the high-frequency modes.
We computed the electron-phonon coupling to be $\lambda_{\mathrm{e-ph}} \approx 0.55-0.65$. This suggests that \ce{LaRh3B2} is a weak-to-moderate coupling superconductor. The McMillan formula was then used to estimate the superconducting critical temperature $T_c$ \cite{McMillan,Carbotte}:
\begin{equation}
    T_c = \frac{\omega_{log}}{1.2} e^{\bigl[\frac{-1.04(1+\lambda)}{\lambda (1-0.62\mu^*) - \mu^*}\bigr]}
\end{equation}
with $\omega_{log}$ being related to the Eliashberg function:
\begin{equation}
    \omega_{log} = e^{\bigl[\frac{2}{\lambda} \int{\frac{d\omega}{\omega} \alpha^2 F(\omega) \log{\omega}} \bigr]}
\end{equation}
While the Coulomb pseudo-potential $\mu^*$ lies in the typical range [0.1 - 0.2], we obtain values for $T_c$ in fair agreement with the experimental results. Specifically, $T_c \approx 2.6$ K is obtained for $\mu^* = 0.17$ (Fig. \ref{fig:th} d).\\

\section{Summary and Discussion:}    
We report on the electronic structure, phonon spectrum and physical properties of a kagome lattice superconductor LaRh$_3$B$_2$.  The structure of LaRh$_3$B$_2$ is built up of kagome planes of Rh stacked along the $c$-axis with La-B planes separating the kagome planes.
The electronic structure contains all features expected for a 2D kagome lattice including a flat band and Dirac bands and van Hove singularities at various positions in the Brillouin zone and in particular near $E_F$. This is qualitatively consistent with the band-structure observed for the $A$\ce{V3Sb5} kagome metals.  In contrast with the $A$\ce{V3Sb5} materials, however, we did not observe signatures of strong electronic correlations in LaRh$_3$B$_2$. The van Hove singularities in the electronic band structure are situated further away from $E_F$ than for $A$\ce{V3Sb5}, which is a fermiological reason why one would not expect charge ordering or density wave like instabilities as reported for $A$\ce{V3Sb5}.

 The superconductivity in LaRh$_3$B$_2$ seems conventional, and there is no experimental evidence for charge density wave instabilities as reported for the $A$\ce{V3Sb5} materials.  The majority contribution to the DOS at $E_F$ in LaRh$_3$B$_2$ derives from Rh $4$d bands.  This suggests that the role of electronic correlations is weakened in LaRh$_3$B$_2$ compared to the family of $A$\ce{V3Sb5} kagome metals, since the Rh $4d$ orbitals are less strongly-correlated than the V $3d$ orbitals. Interestingly, the computed $\lambda_{\mathrm{e-ph}}$ for \ce{LaRh3B2} is in good agreement with experimentally reported values of $\lambda_{\mathrm{e-ph}}$ for the \ce{CsV3Sb5} compound \cite{Zhong} which, together with the phonon dispersions, suggests a similarity in the principal phonon sector between \ce{LaRh3B2}  and $A$\ce{V3Sb5}. It suggests that a central difference between \ce{LaRh3B2} and the by now more established kagome metals must be found in terms of differing electronic correlations, which are lower in strength for \ce{LaRh3B2}. This may explain the absence of CDW in the \ce{LaRh3B2} kagome metal, and points towards phonon-mediated $s$-wave superconductivity.  Given the large number of materials in the $RT_3$B$_2$ and $RT_3$Si$_2$ families, their possibility of Fermiology-tuning around $E_F$ presents an exciting direction for future work. \\

\emph{Acknowledgments.--} We thank the X-ray facility at IISER Mohali.  JS acknowledges UGC-CSIR India for a fellowship.  The phonon-DFT work was supported by the Deutsche Forschungsgemeinschaft (DFG, German Research Foundation) through Project-ID 258499086-SFB 1170 and
by the Würzburg-Dresden Cluster of Excellence on
Complexity and Topology in Quantum Matter-ct.qmat
Project-ID 390858490-EXC 2147. The research leading to these results has received funding from the European Union’s Horizon 2020 research and innovation programme under the Marie Skłodowska-Curie Grant Agreement No. 897276. The authors acknowledge the Gauss
Centre for Supercomputing e.V. for providing computing time on the GCS Supercomputer SuperMUC-NG at Leibniz Supercomputing Centre.

\end{document}